# GEOMETRIC CHARACTERIZATION OF RAT URINARY BLADDER WALL DURING EX-VIVO FILLING USING MICRO-COMPUTED TOMOGRAPHY (MICRO-CT)

**Fatemeh Azari[1], Anne M. Robertson[1,2], Lori A. Birder[3,4]**
[1]Department of Mechanical Engineering and Materials Science, University of Pittsburgh,
3700 O'Hara Street Benedum Hall of Engineering, Pittsburgh, PA 15261.
FAA104@pitt.edu
[2]Department of Bioengineering, University of Pittsburgh,
3700 O'Hara Street Benedum Hall of Engineering, Pittsburgh, PA 15261.
rbertson@pitt.edu
[3]Department of Medicine, University of Pittsburgh,
3550 Terrace St, Pittsburgh, PA 15213.
[4]Department of Pharmacology and Chemical Biology, University of Pittsburgh,
3550 Terrace St, Pittsburgh, PA 15213.
lbirder@pitt.edu

**SUMMARY**

Current mechanical models of the bladder largely idealize the bladder as spherical with uniform thickness. This present study aims to investigate this idealization using micro-CT to generate 3D reconstructed models of rat bladders at 10-20 micrometer resolution in both voided and filled states. Applied to three rat bladders, this approach identifies shape, volume, and thickness variations under different pressures. These results demonstrate the filling/voiding process is far from the idealized spherical inflation/contraction. However, the geometry idealizations may be reasonable in cases where the filled bladder geometry is of importance, such as in studies of growth and remodeling.

## 1 INTRODUCTION

The bladder's functionality is determined by its geometry, wall thickness, and biomechanical properties, all susceptible to impairment due to aging and disease, exemplified by BOO. The importance of BOO for the population is highlighted by the globally escalating prevalence of benign prostatic hyperplasia (BPH), the central contributor to BOO. BPH has shown a pronounced increase of 70.5% from 51.1 million cases in 2000 to 94 million in 2019 [1,2,3] and is a particularly important medical problem for men aged 50-60 years from lower socio-economic backgrounds. BPH induces a spectrum of urinary dysfunctions [4-5] including bladder wall (BW) hypertrophy, changes in bladder dynamics, trabeculation, diverticula, hematuria, and the formation of bladder stones, all of which severely affect bladder compliance during filling and voiding. This complex scenario underscores the necessity for comprehensive research into the BW's evolving properties during BOO progression so that both pharmacological and surgical interventions for BPH can be improved.

Traditionally, the imaging modality of choice for BOO patients has been ultrasound; however, it fails to offer comprehensive insights into wall changes arising from tissue growth and remodeling. Since 1991, there has been intermittent exploration into the geometrical properties of the BW during the filling and voiding cycles. Nonetheless, this field of inquiry is hindered by the paucity of integrated interdisciplinary methodologies and the requisite sophisticated equipment [4,5,6]. Existing scholarly literature delineates two primary experimental approaches for the elucidation of BW properties: analyses conducted at the organ level and those at the tissue strip level [7,8,9]. While investigations at the tissue level, such as uniaxial and biaxial testing, facilitate a detailed examination of layer-specific properties, associated studies of whole bladder structure and function are needed for understanding whole-organ function [10,11,13].

Moreover, conventional approaches to bladder mechanics characterization have historically adopted an oversimplified model of the bladder, conceptualizing it as a uniformly thick, spherical vessel. While these idealizations may be appropriate in some settings, they fail to account for the organ's complex geometry including the changing and non-uniform wall thickness during the filling process. More sophisticated models of the bladder are needed to authentically reproduce the bladder's complex biomechanical behavior. In response to this gap, our investigation employed micro-CT to precisely quantify the geometric properties of the bladder wall within an ex-vivo filling model.

## 2 METHODOLOGIES

### 2.1 Ex-Vivo Filling of the Urinary Bladder Organ

The urinary bladder, ureters, and urethra were surgically removed from three 3-4-month-old female Sprague-Dawley rats and immediately placed in a HEPES-buffered physiological saline solution (HB-PBS) with a composition of 134 mM NaCl, 6 mM KCl, 1 mM $MgCl_2$, 2 mM $MgCl_2$, 10 mM HEPES, and 7 mM glucose, adjusted to a pH of 7.4. Calcium channel blockers were added to the solution to prevent spontaneous contraction of smooth muscle cells (SMC). The ureters were cauterized adjacent to the bladder wall and the urethra tied with 3-0 sutures to a 26G needle. All surrounding connective tissue was excised prior to mounting the bladder in the experimental apparatus. The urethra was cannulated and connected to the syringe pump (BS-8000, Braintree SCIENTIFIC INC) without pre-conditioning, delivering air into the bladder at a syringe translation rate of 1.5 ml/min until reaching a specified transmural pressure of 50-80 mmHg. Transmural pressure was quantitatively measured using a pressure transducer (PX409-OMEGA ENGINEERING INC.) placed near the bladder within the flow circuit. After achieving the target pressure, filling ceased, the valve was closed and the bladder was removed from the apparatus in preparation for micro-CT scanning.

### 2.2 Micro-CT Experimental Design and Morphological Analysis

The central steps to obtain a 3D reconstructed model of each bladder are i) mounting, alignment, and scanning using a high-resolution Skyscan 1272 scanner (Bruker Micro-CT, Kontich, Belgium), ii) 3D reconstruction of the micro-CT Z stacks of 2D images utilizing Nrecon software (Bruker Micro-CT, Kontich, Belgium), iii) morphological analysis of the 3D model using Simpleware ScanIP software (Synopsys, Sunnyvale, California), iv), segmentation of the internal (lumen) and external (ablumen) geometries in Meshmixer software (Autodesk, San Francisco, California), and thickness analysis using Materialize 3-matic software (Materialize GmbH, Munich, Germany). Briefly, the excised bladder was first positioned within a custom-designed holder, ensuring immobilization to prevent data artifacts during the scanning process. After sealing the Luer-lock adapter with parafilm, the holder was mounted in the micro-CT system.

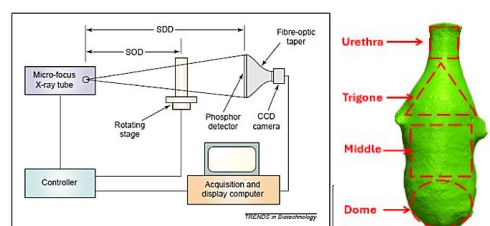

Figure 1: Micro-CT visualization of bladder's internal structure (left) [13], highlighting urethra, trigone, mid-bladder, and dome regions (right)

Micro-CT scans were performed using 80 kV source voltage and 125 $\mu A$ source current. Images were captured at a 10.8 $\mu m$ pixel size using a rotation step of 0.6 degrees, a 2048 x 2048 frame size without filtering, and an exposure time of 400 ms. The reconstruction of these images with NRecon software involved smoothing at level 1, addressing ring artifacts at 50%, and correcting for 2% beam hardening. Scanning time for bladders was kept less than 10 minutes to avoid dehydration. The segmentation process began by thresholding grayscale values to create masks, transforming the reconstructed 2D images into 3D models. These models were then converted into surface models to produce stereolithography (STL) data. The finalized STL files were analyzed in Materialize 3-matics, using the midplane thickness tool to assess wall thickness. Our analysis provided key statistics such as median, average, and standard deviation of wall thickness provided as the output of thickness analysis from 3-Matics. Additionally, a histogram with an adjustable range was generated, ensuring no data points were overlooked. Morphology was compared between voided and filled states with transmural pressures for the filled state of P=50 mmHg, 57 mmHg and 80 mmHg for bladders A, B, C, respectively.

## 3 RESULTS AND CONCLUSIONS

A quantitative assessment of bladder geometry was obtained for three rat bladder specimens, labeled as Bladder A, B, and C, Figure 2. Data was obtained at two inflation states: (1) voided (non-distended state – harvested condition) and (2) filled (distended). Bladder specific morphology results are given in Table 1.

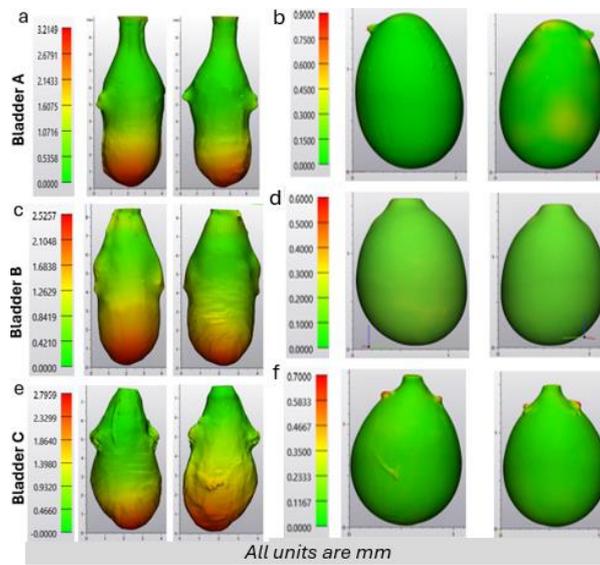

Figure 2: BW thickness at two inflation states, n=3. All units are mm.

As previously noted, the lumen volume undergoes a large-scale change between the voided and filled states. For the three bladders, volume increased from a voided average of 13.63 mm³ $\pm$ 5.50 mm³ (Fig.2 a, c, e) to a filled volume of 850.03 mm³ $\pm$ 123.63 mm³ (Fig.2 b, d, f). On average, the bladder volume increased 70-fold between the voided and filled state. The bladder specific increase was 38-fold (Bladder A), 91-fold (Bladder B) and 80-fold (Bladder C).

| Bladder Volume ($mm^3$) | | |
|---|---|---|
| | Voided State | Inflated State |
| Bladder A | 21.30 | 835.70 |
| Bladder B | 10.91 | 1008.10 |
| Bladder C | 8.69 | 706.29 |
| *Average* | 13.63 | 850.03 |
| *STD* | 5.50 | 123.63 |
| Bladder Median Thickness (mm) | | |
| | Voided State | Inflated State |
| Bladder A | 0.45 | 0.13 |
| Bladder B | 0.77 | 0.09 |
| Bladder C | 0.73 | 0.08 |
| *Average* | 0.65 | 0.10 |
| *STD* | 0.14 | 0.02 |

Table 1: Summary of whole bladders volumes and Thickness

In the voided state, the wall thickness was highly non-uniform. For all three cases, the BW for the voided state was markedly thicker at the dome compared to the mid-bladder and trigonal areas. In particular, the average maximum dome thickness for all three bladders was 2.84 mm ±0.28 mm while the median thickness of the middle and trigonal areas was 0.65 ± 0.14, respectively. In contrast, upon distension (Fig. 2, b, d, f), all three bladders have a relatively uniform, thin wall with an average median thickness of 0.10mm ±0.02mm.

Numerous models, including our own [3], have idealized the bladder with a spherical configuration and homogenous wall thickness [3,6,10], an approximation advantageous for deriving analytical solutions. The current work suggests this may be a reasonable approximation in studies where the filled bladder geometry is of importance. For example, in a recent study of growth and remodeling for BOO bladders, the filled (spherical) bladder was used as the reference configuration for defining the homeostatic stretch of both collagen fibers and smooth muscle cells [3]. However, it is possible that even the full bladder will show deviations in shape and wall thickness from this idealization in pathological states or as a result of aging [13]. A comparison of the shape and wall thickness between the voided and filled states, Fig 2, demonstrates that even for the healthy bladder, the filling process varies spatially over the wall of the bladder and is not well represented by simple spherical inflation. It is anticipated that heterogeneities in wall thickness will be even more complex during disease and a result of aging [13]. Hence, the changing shape and wall thickness during micturition will need to be quantified in these cases.

In summary, this work highlights the necessity to employ sophisticated imaging modalities, such as high-resolution micro-CT, to reveal bladder morphology and identify region-specific alterations in wall thickness throughout the filling and voiding process. Such high-resolution data is vital for computational mechanics models of the bladder needed for studying evolving bladder functionality during diseases such as bladder outlet obstruction (BOO). The authors express their gratitude for the funding received from NIH-R01 AG056944 and NIH-R01 DK133434.